\documentclass[floatfix,twocolumn,a5paper,aps,prc,superscriptaddress,showpacs]{revtex4}
\usepackage{graphicx}
\usepackage{amsmath}
\usepackage{amsfonts}
\usepackage{dcolumn}
\newcommand\mc[1]{\multicolumn{1}{c}{$#1$}}
\newcolumntype{e}[1]{D{.}{\cdot}{#1}}

\begin{document}

\title {A simple model for the quenching of pairing correlations effects\\ in
  rigidly deformed rotational bands}
\date{\today}

\author{P. Quentin}
\affiliation{Centre d'\'Etudes Nucl\'eaires de Bordeaux-Gradignan
  (CNRS-IN2P3 and Universit\'e Bordeaux I), Le Haut Vigneau BP 120, F-33175 
  Gradignan, France}
\affiliation{Theoretical Division, T-DO (Los Alamos National Laboratory),
  PO Box 1663, Los Alamos, NM 87545, USA}
\author{H. Lafchiev}
\affiliation{Centre d'\'Etudes Nucl\'eaires de Bordeaux-Gradignan
  (CNRS-IN2P3 and Universit\'e Bordeaux I), Le Haut Vigneau BP 120, F-33175 
  Gradignan, France}
\affiliation{Institute of Nuclear Research and Nuclear Energy (Bulgarian
  Academy of Sciences), Tzarigradsko Chaussee 72, 1784 Sofia, Bulgaria}
\author{D. Sams\oe{}n}
\email{samsoen@cenbg.in2p3.fr}
\affiliation{Centre d'\'Etudes Nucl\'eaires de Bordeaux-Gradignan
  (CNRS-IN2P3 and Universit\'e Bordeaux I), Le Haut Vigneau BP 120, F-33175 
  Gradignan, France}
\author{I. N. Mikhailov}
\affiliation{Bogoliubov Laboratory of Theoretical Physics
  (Joint Institute of Nuclear Research), Joliot-Curie str. 6, 141980
  Dubna (Moscow Region), Russia}
\affiliation{Centre de spectrom\'etrie Nucl\'eaire et de Spectrom\'etrie
  de Masse, (CNRS-IN2P3 and Universit\'e Paris XI), B\^at. 104, 
  91406 Orsay-Campus, France}

\begin{abstract}
Using Chandrasekhar's \textit{S}-type coupling between rotational and
  intrinsic vortical modes one may simply reproduce the HFB dynamical
  properties of rotating nuclei within Routhian HF calculations free of
  pairing correlations yet constrained on the relevant so-called Kelvin
  circulation operator.
From the analogy between magnetic and rotating systems, one derives a model
  for the quenching of pairing correlations with rotation, introducing a
  critical angular velocity---analogous to the critical field in
  supraconductors---above which pairing vanishes.
Taking stock of this usual model, it is then shown that the characteristic
  behavior of the vortical mode angular velocity as a function of the global
  rotation angular velocity can be modelised by a simple two parameter
  formula, both parameters being completely determined from properties of the
  band-head (zero-spin) HFB solution.
From calculation in five nuclei, the validity of this modelised Routhian
  approach is assessed.
It is clearly shown to be very good in cases where the evolution of rotational 
  properties is only governed by the coupling between the global rotation and
  the pairing-induced intrinsic vortical currents.
It therefore provides a sound ground base for evaluating the importance of
 coupling of rotation with other modes (shape distortions, quasiparticle
 degrees of freedom).
\end{abstract}
\maketitle

\section{Introduction}
\label{sec:intro}
In a previous paper (hereafter referred as I) we have provided some
evidence that the dynamical effects of pairing correlations at finite spins
could be very well represented indeed by an intrinsic flow being both
non-deforming and counter-rotating (with respect to the global rotation)
\cite{LSQ03:equiv}. This has been borne out from the study of rotational bands
in three heavy nuclei chosen to represent rather different cases (as far as
nucleon numbers, deformation, spin values or pairing correlation content are
concerned). To achieve this, we have performed two sets of microscopic
calculations under a Routhian-type constraint (namely including in the
variational quantity for the rotating case a $-\vec\Omega\cdot\vec{J}$ term
with usual notation).
The first calculations have included pairing correlations within the
Hartree-Fock-Bogoliubov (HFB) formalism. The second ones were of the
Hartree-Fock type (i.e., without any pairing correlations) with a double
Routhian-type constraint, as presented in Refs. 
\cite{MQS97:vort,SQB99:HF,SQM99:period}
and recalled here in Appendix \ref{app:inertia}, which we will refer to below
as HF+V calculations .
The ``measuring stick'' for the latter was the so-called Kelvin circulation
operator $\hat{K}_1$ (see, e.g., Ref. \cite{Ros92}) which is well suited for
the description of collective modes dubbed after Chandrasekhar as 
\textit{S}-type ellipsoids \cite{Chandra}.
Indeed, upon imposing via the Routhian double-constraint that the second type
of solutions should have the same $\langle\hat{K}_1\rangle$ expectation
value at a given value of the spin $I$ as those corresponding to the
HFB calculations, it has been demonstrated in I that both calculations yielded
the same rotational properties.

As shown also in I, the intensity of such counter-rotating currents (measured,
e.g., by $\langle\hat{K}_1\rangle$) behaves as a function of the global
rotation angular velocity $\Omega$ in a very characteristic way.
One may figure out that 
such a reactive mode
of the fluid should be proportional to the excitation velocity field intensity
(i.e., $\Omega$).
Moreover it is also safe to assume that $|\langle\hat{K}_1\rangle|$ should be
an increasing function of the pairing correlations measured by some power of
$-E_{\mathrm{corr}}$ where $E_{\mathrm{corr}}$ is the (negative) pair
correlation energy. Now, it is well known (see Ref. \cite{MV60}) that pairing
correlations tend
to decrease upon increasing $\Omega$.
When $\Omega$ gets larger therefore, one should expect a balance between two
competing phenomena resulting in a maximum of $|\langle\hat{K}_1\rangle|$ as a
function of $\Omega$ between two limiting cases.
One is $\Omega = 0$ and the other is obtained for a critical value $\Omega_c$
corresponding to the alleged transition between a ``superfluid'' and a
``normal'' phase of nuclear matter.

A word of caution is worth adding at this point. It is absolutely not clear
that a transition between normal and superfluid nuclear phases should occur
within a physically acceptable range of $\Omega$ values for such finite
Coulombian systems as atomic nuclei.
In this respect it is well known that the usual Bogoliubov treatment (or its
Bogoliubov-Valatin or BCS reduction) tends to overestimate the instability of
the superfluid solution in low pairing regimes in general, and at high
$\Omega$ values in particular.
Only approaches which would conserve explicitly the particle number, could
cast a priori some light on the existence of such a transition. Such
calculations are currently achieved \cite{Routh:htda} within an extension of
the approach developed in Ref. \cite{PQL02:htda} to describe correlated states
in even-even nuclei (there, of course, the time-reversal symmetry was not
broken).
In the present paper however, we will overlook this problem and stick to the
usual HFB approach which entails naturally that this particular part of our
description in the limit of high $\Omega$ regime (i.e., close to the critical
$\Omega_c$ value) should be considered as rather schematic indeed.

\begin{figure}[h]
  \centerline{%
    \includegraphics[scale=1.7]{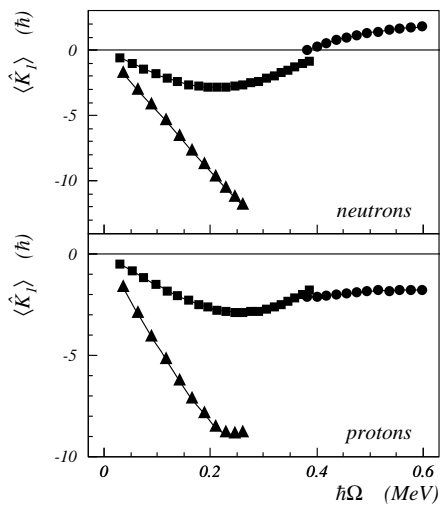}}
  \caption{Kelvin circulation mean value in $\hbar$ units for neutrons (upper
    panel) and protons (lower panel) in the three rotational bands studied in
    I. The convention in use is the following: circles for $^{150}$Gd,
    squares for $^{192}$Hg and triangles for $^{254}$No. Calculations are
    performed within the HFB formalism for $^{254}$No and
    within the HFB+LN formalism for $^{150}$Gd and $^{192}$Hg.}
  \label{fig:kelvin}
\end{figure}

The above described competition pattern has been illustrated in I by curves
(see Fig.~1 there) exhibiting $\langle\hat{K}_1\rangle$ as a function of
$\Omega$. The main features of these results are sketched here in
Fig.~\ref{fig:kelvin} for the isoscalar values of $\langle\hat{K}_1\rangle$ in
the three considered rotational bands (yrast superdeformed bands of $^{150}$Gd
and $^{192}$Hg, and ground-state band of $^{254}$No). As seen on this
Figure, it is found only in the case of the $^{192}$Hg nucleus that the range
of $\Omega$ values spanned by our calculations (or more precisely by
corresponding available experimental data) allows to exhibit a clean cut
extremum in the corresponding curve.
For the $^{150}$Gd nucleus one merely displays the up-going part of the curve
(i.e., after the minimum of $\langle\hat{K}_1\rangle$ and before $\Omega_c$)
while, quite on the contrary, for the $^{254}$No nucleus, one only sees in
Fig.~\ref{fig:kelvin} the down-going part of the curve (i.e., just above
$\Omega = 0$) reaching its minimum, only for protons.

In addition, in the $^{150}$Gd case, a very interesting phenomenon is present
and is worth discussing. As presented in I, one sees that substantial
$\langle\hat{K}_1\rangle$ values are present for both neutrons and protons
(namely about $+1\;\hbar$ and $-2\;\hbar$ respectively) even when pairing
correlations have disappeared around $\hbar\Omega=0.6$~MeV. 
This phenomenon, also present for the proton Kelvin circulation
in $^{192}$Hg (whose pairing correlations vanish around
$\hbar\Omega=0.4$~MeV), is clearly a consequence of shell effects as it has
been shown in I to be also present within purely rotating Routhian HF
calculations.
The behavior of $\langle\hat K_1\rangle$ in $^{254}$No is also worth a closer
look. Indeed, as the neutron pairing energy is almost vanishing (see Fig.~2 in
I) even for $\hbar\Omega=0$, a large negative Kelvin circulation mean
value is observed in Fig.~\ref{fig:kelvin} for neutrons.
This phenomenon also could be attributed partly to shell effects, but it is
probably mostly related to a dragging effect of the protons.

\begin{figure}[h]
  \centerline{%
    \includegraphics[scale=1.7]{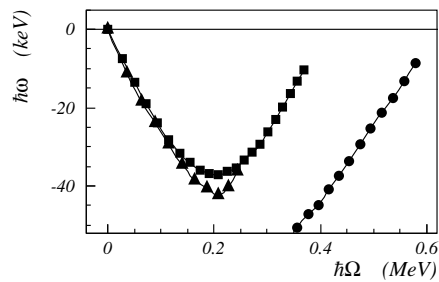}}
  \caption{Kelvin circulation velocity $\omega$ (in keV) as a function of the
    rotation angular velocity $\Omega$ in MeV for the three nuclei considered
    in I and calculated within the constrained HF+V formalism. The symbol
    conventions are the same as those in Fig.~\ref{fig:kelvin}}
  \label{fig:velocity}
\end{figure}

Another way of representing the same trends is to plot, as done here in
Fig.~\ref{fig:velocity} for the three considered nuclei, the values of the
Lagrange multiplier $\omega(\Omega)$ of the Routhian (\ref{eq:routh})
determined in I to obtain the relevant HFB $\langle\hat{K}_1\rangle$ values as
a function of $\Omega$.
As seen here they exhibit a parabolic-like behavior. Being related
explicitly to the extra-contribution to $\langle\hat{K}_1\rangle$ due to
pairing correlations, it is not surprising that such $\omega(\Omega)$ curves
are all located in our case in the $\omega < 0$ part of the plane.

It is the aim of the present paper to understand quantitatively the pattern of
these $\omega(\Omega)$ curves. To put it in more operational terms we want to
find whether we are able to deduce from the geometrical and pairing
correlation properties of HFB solutions at zero-spin, the properties of
rotational states. This of course would be only possible in cases
where one deals with a ``good'' rotational band 
behavior, namely wherever the rotation does not couple significantly with
other degrees of freedom like vibrations (e.g., no rotational stretching or
anti-stretching) or single-particle excitations (e.g., no back-bending).

The paper will be organized as follows.
In Section \ref{sec:analog} we will briefly develop a simple magnetic-like
model inspired from Ref. \cite{MV60} using the well-known supraconductor
critical field concept (see, e.g., \cite{Kittel:solids}) to describe the
reduction of pairing correlations upon increasing $\Omega$.
Section \ref{sec:assum} will be devoted to
the proposition and discussion of an ansatz for the relation between $\omega$
and $\Omega$.
We will present in Section \ref{sec:params}, the prescriptions to determine
the parameters of the above relation from the zero-spin HFB solution of a
given nucleus.
Finally, the validity of the model so established will be tested against
actual HFB Routhian calculations for five nuclei (including the three already
considered in I) and some conclusions will be drawn in Section
\ref{sec:conclu}.

\section{A simple magnetic-like analog}
\label{sec:analog}

In the simple model presented here (which is directly inspired from the
rotational quenching mechanism of pairing correlations proposed by Mottelson
and Valatin \cite{MV60}, the so-called Coriolis anti-pairing effect), we
assume that the action of a global
rotation on a pair of particles moving in opposite directions on a given
orbital may be described roughly in terms of two interacting magnetic dipoles
($\vec{D}_1$, $\vec{D}_2$) plunged in an external magnetic field $\vec{B}$.
The total energy of such a system is made of a term $E_{\mathrm{corr}}$
describing the dipole-dipole coupling and a term $E_m$ taking into account the
coupling with the external field. However, while the latter is understood as a
magnetic field-dipole interaction term (see, e.g., Ref. \cite{Jackson},
Chapter 5)
\begin{equation}
  E_{\mathrm{m}} = -\beta \vec{B} \cdot (\vec{D}_1 + \vec{D}_2) \quad ; \quad
  \beta > 0\,, 
\label{eq:coupling}
\end{equation}
the former does not originate, in our case, from a magnetic dipole-dipole
interaction (which actually would tend to align $\vec{D}_1$ and $\vec{D}_2$),
but rather from the strong interaction favoring quite on the contrary an
anti-alignment of $\vec{D}_1$ and $\vec{D}_2$.
Therefore $E_{\mathrm{corr}}$ (which will be later called the correlation
energy) is not given by the standard magnetic dipole-dipole expression (see,
e.g., Ref.  \cite{Jackson}, Chapter 4) but rather, assuming the simplest
possible form compatible with the physical requirements, as:
\begin{equation}
\label{A2}
E_{\mathrm{corr}} =  \alpha \big( \vec{D}_1 \cdot\vec{D}_2 - |\vec{D}_1|
|\vec{D}_2| \big)\quad ; \quad \alpha > 0 \,.
\end{equation}

Assuming $\vec{B}$ (whose norm will be denoted as $B$) to lie on the 3-axis
and $\vec{D}_1$ to belong to the (1,3) plane, one gets for the total energy
the following expression (where $\theta_i$, $\phi_i$ are the usual spherical
coordinates determining the $\vec{D}_i$ direction):
\begin{equation}
\label{A3}
\begin{aligned}
  E_{\mathrm{tot}} &= \alpha d^2 \big[\sin(\theta_1)\sin(\theta_2)\cos(\phi_2)
  +\cos(\theta_1)\cos(\theta_2)-1\big]\\
  &- \beta B d \big[\cos(\theta_1) + \cos(\theta_2)\big]
 \end{aligned}
\end{equation}
where $d$ is the common norm of $\vec{D}_1$ and $\vec{D}_2$.

The variation of $E_{\mathrm{tot}}$ with respect to $\phi_2$ leads to
an extremum either at $\sin\phi_2=0$ or else at $\sin\theta_1$ and/or
$\sin\theta_2=0$. In the latter case, it is easy to check that to get
an extremal energy, the vanishing of $\sin\theta_1$ entails the
vanishing of $\sin\theta_2$ and conversely. One is thus left with $\sin\theta_1
= \sin\theta_2=0$.

Discarding for a while this solution let us concentrate now on the
former case. Clearly to get a minimal energy $\phi_2=\pi$ is to be
preferred to $\phi_2=0$. The variation of $E_{\mathrm{tot}}$ with respect to
$\theta_1$ and $\theta_2$ leads now to
\begin{equation}
  \beta B \sin\theta_1 = \beta B\sin\theta_2 = \alpha d \sin(\theta_1+\theta_2)
\end{equation}
which admits two classes of solutions:
\begin{subequations}
  \begin{align}
    \theta_1=\theta_2&=\theta \,,\\
    \theta_1+\theta_2&=\pi \,.
  \end{align}
\end{subequations}
In the first case, the remaining variational equation
\begin{equation}
  \beta B\sin\theta= \alpha d \sin(2\theta)
\end{equation}
has two solutions. The first one is defined by
\begin{equation}
  \cos\theta = \frac{\beta B}{2\alpha d}
  \label{coscritic}
\end{equation}
whose corresponding equilibrium energy is
\begin{equation}
   E_{\mathrm{tot}} = -\alpha d^2 - \frac{\beta^2B^2}{2\alpha} \,.
  \label{totcritic}
\end{equation}
The second solution is defined by $\sin\theta=0$. Therefore, all
variational solutions other than the one of eq. (\ref{coscritic})
correspond to  $\sin\theta_1 = \sin\theta_2 = 0$, namely :
\begin{subequations}
    \label{allothers}
    \begin{align}
     \theta_1=\theta_2=0 \quad&;\quad E_{\mathrm{tot}} =  
           -2\beta Bd\,, \label{other:a}\\
     \theta_1=\theta_2=\pi \quad&;\quad E_{\mathrm{tot}} = 
           2\beta Bd\,, \label{other:b}\\
     \bigg\{\begin{array}{l}
       \theta_1=0,\:\theta_2=\pi \\
       \theta_1=\pi,\:\theta_2=0
     \end{array}\;
     &;\quad E_{\mathrm{tot}} = -2\alpha d^2 \,. \label{other:c}
    \end{align}
\end{subequations}
While searching for the most stable equilibrium solution, it is found
upon comparing the total energies given in eqs. (\ref{totcritic}) and
(\ref{other:a}-\ref{other:c}) that the former is to be retained (since
in particular $\alpha d^2 + \beta^2B^2/(2\alpha) - 2\beta Bd$ is clearly
positive).

For the solution of eq. (\ref{totcritic}), as expected, there is an
upper limit for the norm $B_c$ of $\vec B$ given by
\begin{equation}
\label{A6}
B_c = \frac{2\alpha d}{\beta}\end{equation}
corresponding to a full alignment ($\theta = 0$). The corresponding
correlation energy $E_{\mathrm{corr}}$ at equilibrium assumes the following
form:
\begin{equation}
\label{A8}
E_{\mathrm{corr}} = 2\alpha d^2 \left[\left(\frac{B}{B_c}\right)^2 - 1\right]
\,. 
\end{equation}
One may note, parenthetically, that the critical field value is obtained upon
equating twice the correlation energy without
field with the dipole magnetic interaction in the aligned case
\begin{equation}
\label{A9}
-4 \alpha d^2 = - \beta B d
\end{equation}
which is very similar in spirit with the way in which one determines
the critical magnetic field in the classical theory of
supraconductivity (see, e.g., Ref. \cite{Kittel:solids}).

\section{Model assumptions in the case of nuclear rotational states}
\label{sec:assum}

As well known, the equations of motion of a particle coupled to a magnetic
field by its charge, are similar to those of a massive particle moving in a
rotating (non-inertial) frame upon identifying, with a proportionality
constant, the magnetic field $\vec{B}$ and the angular velocity
$\vec{\Omega}$. It allows therefore to schematically model the coupling of a
single-particle orbital motion in the rotating frame with the global rotation
(in the laboratory frame) as the magnetic dipole interaction considered in the
previous section [see eq. (\ref{eq:coupling})]. Similarly, the corresponding
interaction energy between two particles moving in opposite directions on a
same orbit could be approximated as:
\begin{equation}
  E_{\mathrm{corr}} = \lambda \left[\left(\frac{\Omega}{\Omega_c}\right)^2 -
    1\right] \,, 
\label{eq:pairing}
\end{equation}
introducing thus a critical angular velocity $\Omega_c$ which will be defined
below.

The preceding could pertain, a priori, to the description of the internal and
external energies of a single pair. It is our contention, however, that such a
quantity behaves, up to a multiplicative constant, as the energies of all the
pairs to be considered in the nucleus. Therefore, after suitably redefining
the phenomenological parameter $\lambda$, one may consider the expression
above given in eq.~ (\ref{eq:pairing}) as the total nuclear pairing correlation
energy.

Now, we are proceeding with the major assumption behind our model. Namely, we
assume that the vector $\vec{\omega}$ associated with the intrinsic
vortical motion is proportional to the global rotation angular velocity
$\vec{\Omega}$. The corresponding proportionality factor is negative and we
assume that its absolute value is an increasing function of the total
correlation energy of eq. (\ref{eq:pairing}). Specifically, we propose that the
projection $\omega$ of the vector $\vec{\omega}$ on the vector $\vec\Omega$ is
given by:
\begin{equation}
  \omega = - \alpha (E_{\mathrm{corr}})^{\gamma} \Omega 
  \quad ; \quad \alpha,\gamma >  0\,.
\label{A13}
\end{equation}
Upon inserting in such an ansatz, the correlation energy expression of
eq. (\ref{eq:pairing}), one gets for $\omega$
\begin{equation}
  \omega = - k \Omega \left[1 -
    \Big(\frac{\Omega}{\Omega_c}\Big)^2\right]^{\gamma} \quad ; \quad k > 0
  \mathrm{\;and\;}  \Omega \in{[0,\Omega_c]}\,.
\label{eq:omega}
\end{equation}

For $\gamma \neq 1$, one gets two extrema for $\omega$ corresponding to:
\begin{equation}
\label{A15}
\Omega_1 = \frac{\Omega_c}{\sqrt{1 + 2 \gamma}} \quad ; \quad \Omega_2 =
\Omega_c 
\end{equation}
(for $\gamma = 1$ one would have only the variational solution corresponding
to $\Omega_1$).
The correlation energy associated with the $\Omega_1$ 
solution, expressed in units of its value at zero angular momentum ($\Omega =
0$) is:
\begin{equation}
  \frac{E_{\mathrm{corr}}(\Omega_1)}{E_{\mathrm{corr}}(0)} = 
  \frac{2 \gamma}{1 + 2 \gamma}\,.
\label{eq:ecorr}
\end{equation}

\section{Determination of the model parameters}
\label{sec:params}

Let us summarize the modeling which has been performed so far. With eq.
(\ref{eq:omega}), we have a three-parameter expression for the intrinsic
vorticity parameter $\omega$ in terms of the global rotation angular velocity
$\Omega$, namely:
\begin{enumerate}
\item[i)] a scale parameter ($\Omega_c$) for the abscissa variable $\Omega$,
\item[ii)] a scale parameter ($k$) for the studied quantity $\omega$, and
\item[iii)] a power parameter $(\gamma)$ expressing the nature of the pairing
  correlation 
  dependence of $\omega$.
\end{enumerate}

The former parameter $\Omega_c$ may be fixed a priori for each considered
nucleus as sketched before.  In eq. (\ref{A9}), it is given upon equating
twice the correlation energy without field (thus, here, without rotation),
with the equivalent of the dipole-magnetic coupling energy in the aligned
case, namely the rotational energy without pairing. We therefore get 
\begin{equation}
\Omega_c = \sqrt{\frac{4 E_{\mathrm{corr}}(\Omega = 0)}
                      {\mathfrak{J}_{\mathrm{rig}}}}\,.
\label{omegac-1}
\end{equation}
In the above $\mathfrak{J}_{\mathrm{rig}}$ represents the moment of inertia of
an unpaired nucleus which could be reasonably well represented by the
rigid-body moment of inertia associated with a one-body reduced local density
$\rho(\vec{r})$ obtained, e.g., in an Hartree-Fock or Hartree-Fock-Bogoliubov
calculation as:
\begin{equation}
\mathfrak{J}_{\mathrm{rig}} = m \int \rho(\vec{r})
\left(\vec{r} - (\vec u \cdot\vec r)\vec u \right)^2 \; {\rm d}^3r\,,
\label{eq:rigid}
\end{equation}
where $\vec u$ is the unit vector $\vec{\Omega}/\Omega$.
However, in doing so one would introduce an unwanted contribution of
shell effects and/or pairing correlations to $\mathfrak{J}_{\mathrm{rig}}$
through the density function $\rho(\vec{r})$.
Even though the effect of the latter is generally rather small we propose
here, to remain in the spirit of the model (associated with bulk nuclear
properties), to rather consider the self-consistent semiclassical moment of
inertia $\mathfrak{J}_{\mathrm{ETF}}$ (within the
Extended Thomas-Fermi scheme) of Ref. \cite{BBQ94} using the same effective
interaction at the relevant deformation.

In most current Routhian HFB calculations (including
ours \cite{Laf01:nobel}) using the Skyrme interaction for the ``normal'' mean
field, there is an inconsistency in the interaction used in the Hartree-Fock
(``normal'') part and in the part dealing with pairing correlations.
As a result, the naturally defined correlation energy, namely the energy
difference between the correlated and the non-correlated solutions, is not
relevant. Thus we must replace in eq. (\ref{omegac-1}), the correlation energy
$E_{\mathrm{corr}}(\Omega=0)$ by an unequivocal quantity. 
The ``abnormal'' part of the total HFB energy---sometimes
called (see, e.g., Ref. \cite{FQ73:chf} and Appendix \ref{app:forschle}) the
``pair condensation energy'',
$E_{\mathrm{cond}}$---could be considered in practice in this case, as a more
reliable index of the amount of pairing correlations.
As demonstrated in model calculations performed 
in the three cases considered in I according 
to the approach of Ref. \cite{PQL02:htda} which is free from 
the usual HFB breaking of the particle number symmetry, the pair 
condensation energy is found to be roughly equal to twice the pair 
correlation energy (see appendix \ref{app:forschle} for details). The equation 
(\ref{omegac-1}) yielding  the critical value $\Omega_c$ 
becomes thus
\begin{equation}
\Omega_c = \sqrt{\frac{2E_{\mathrm{cond}}(\Omega=0)}
                      {\mathfrak{J}_{\mathrm{ETF}}}}\,.
\label{omegac}
\end{equation}

The scale parameter $k$ can be determined in the following way. If we assume
that our relation $\omega(\Omega)$ is of any global relevance, it should also
be valid in the low angular velocity regime. Namely, one should have for
$\Omega\ll\Omega_c$ whatever the choice of the power $\gamma$:
\begin{equation}
k=-\frac{\omega}{\Omega}\,.
\end{equation}
In this adiabatic regime, one obtains an expression for the total
collective kinetic energy in a HF+V approach which is quadratic in $\Omega$
and $\omega$, as with the notation of Ref. \cite{MQ95:stagg}
\begin{equation}
  E(\Omega, \omega)=\frac{1}{2}
  \left( A \omega^2 + 2 B \omega \Omega + C \Omega^2 \right)\,.
  \label{eq:quadratic}
\end{equation}
Upon including the above limiting expression for the ratio of $\omega$ and
$\Omega$ one gets the expectation value of the total angular momentum as
\begin{equation}
  \langle\hat J_1\rangle = \frac{\partial\: E(\Omega, \omega)}{\partial
    \Omega} \Big|_{\omega=-k\Omega}= (C - k B)\: \Omega\,,
\end{equation}
which in turn yields the value of the dynamical moment of inertia within our
model:
\begin{equation}
  \mathfrak{J}^{(2)}_{\mathrm{mod}} = \frac{\mathrm{d} \:\langle\hat
    J_1\rangle} {\mathrm{d} \Omega} = C - k B\,
  \label{eq:inertiamodel}
\end{equation}
as shown in Appendix \ref{app:inertia}.

Now, we define the HFB Routhian dynamical moment of inertia
$\mathfrak{J}_{\mathrm{HFB}}$ (in the low angular velocity regime) and require
it has the same value as $\mathfrak{J}^{(2)}_{\mathrm{mod}}$:
\begin{equation}
  \mathfrak{J}_{\mathrm{HFB}} = \frac{\langle\hat J_1\rangle}{\Omega} =
  \mathfrak{J}^{(2)}_{\mathrm{mod}}\,. 
\end{equation}
One is left therefore, now, with the problem of determining the 
moments $A$, $B$ and $C$. A natural way to do so is of course
to calculate the total energy surface as a function of $\Omega$ 
and $\omega$ in the vicinity of zero for both parameters
through doubly constrained HF calculations (called in I 
Routhian HF + V calculations) and then
perform a quadratic fit. This has been done indeed 
for the five considered cases (see Appendix \ref{app:quadratic} for the
results). Even though this does not amount to performing very much 
time-consuming calculations, we have developed another approach 
which is much simpler and seems to be quite sufficient for the model
calculations tested here.

One may evaluate from the single-particle eigenstates of the corresponding
Hartree-Fock solution, the mass parameters $A$, $B$ and $C$ through the Inglis
cranking formula \cite{Ing54,Ing56} where the cranking operators are (see
Ref. \cite{MQ95:stagg}) the first components of the total angular momentum
$\hat J_1$ and of the Kelvin circulation operator $\hat K_1$. With usual
notation one has thus
\begin{equation}
  \begin{aligned}
    C = 2 \sum_{p,h} \frac{|\langle p|\hat
      J_1|h\rangle|^2}{\epsilon_p-\epsilon_h} \,,\\
    B = 2 \sum_{p,h} \frac{\langle p|\hat J_1|h\rangle \langle h|\hat
      K_1|p\rangle}
    {\epsilon_p-\epsilon_h}\,,\\
    A = 2 \sum_{p,h} \frac{|\langle p|\hat
      K_1|h\rangle|^2}{\epsilon_p-\epsilon_h}\,.
\end{aligned}
\label{eq:inglis}
\end{equation}

However it is well known (see e.g. Ref. \cite{GQ80}) that one should not
compare dynamical moments of inertia obtained in a Routhian approach with
comparable moments in an Inglis cranking approach. Indeed the latter lack
the so-called Thouless-Valatin \cite{TV62} terms which are merely coming from
the time-odd density response to the self-consistent time-odd Hartree-Fock
mean field. For instance, it has been estimated in Ref. \cite{LGD99}
for the usual cranking moment $C$ and in the presence of pairing correlations
that one gets with the Inglis cranking formula an under-estimation of the
Routhian value of about $\tfrac{1}{3}$. In view of the similarity of the two
operators $\hat J_1$ and $\hat K_1$ it is reasonable to further assume that
the same multiplying factor should be applied also for the mass parameters $A$
and $B$. These approximations and the value of the Thouless-Valatin correction
parameter $\eta$ [such that the actual values should be enhanced from 
their Inglis cranking crude estimates by a factor $(1+\eta)$] have been 
assessed from doubly constrained HF calculations in the five considered 
cases in Appendix \ref{app:quadratic}. As we will see in the next section, two
different Skyrme effective interactions have been used in our calculations
(SIII \cite{BFG75:SIII} and SkM$^*$ \cite{Bar82:SkM}). The corresponding
values for $\eta$ are 0.2 and 0.1 respectively.

The difference between these values as well as with the value 
proposed in Ref. \cite{LGD99} may have two origins. The first is
that the interactions used in the various self-consistent calculations are 
different.
As shown in \cite{GQ80b}, the Thouless-Valatin corrective terms for Skyrme
interactions increase when the difference between the real nucleon mass ($m$)
and its effective value in nuclear matter ($m^*$) increases.
If this evidence for Skyrme interactions is to be extended to the Gogny force
in use in \cite{LGD99}, it appears that at least a part of this difference may
be due to the fact that the ratio $m^*/m$ equals 0.70 for the considered Gogny force,
0.75 for the SIII force, and 0.79 for the SkM$^*$ force.
The second is that in Ref. \cite{LGD99} one has performed Routhian calculation
\textit{with} pairing correlations.

As a result one gets the following value for $k$:
\begin{equation}
  k = \frac{C - (1 - \eta) \mathfrak{J}_{\mathrm{HFB}}}{B}\,.
  \label{eq:k}
\end{equation}

The last parameter whose value has to be pinpointed is the power $\gamma$ of
the correlation energy dependence of $\omega$. By definition if $\gamma=1$
the parameter $\omega$ depends linearly on the pairing energy (correlation or
condensation energies) whereas if $\gamma=0.5$ it would depend linearly on
the average pairing gap. We have found no binding arguments to decide a priori
on this value. Besides, as seen from equation (\ref{eq:ecorr}) in the $\gamma=$
$0.5$--$1$ range the location of the maximum value of $|\omega|$ is not
violently contingent upon a specific figure for $\gamma$. We have therefore
decided not to use the freedom of fitting this parameter and therefore we have
somewhat arbitrarily decided to take $\gamma=1$.

\section{Results and Conclusions}
\label{sec:conclu}

\begin{figure}[ht]
  \centerline{%
    \includegraphics[scale=1.7]{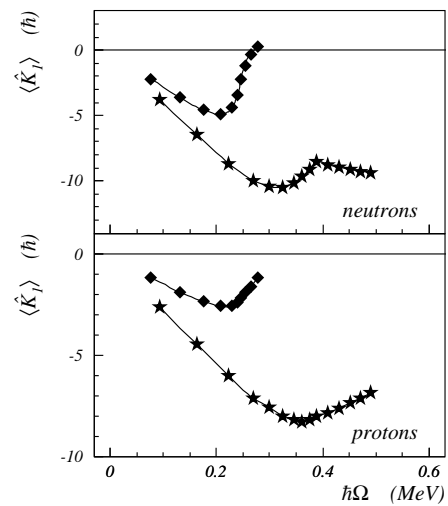}}
  \caption{Same as Fig. \ref{fig:kelvin} for $^{154}$Sm (diamonds) and
    $^{178}$Hf (stars).}
  \label{fig:morekelvin}
\end{figure}

\begin{figure}[ht]
  \centerline{%
    \includegraphics[scale=1.7]{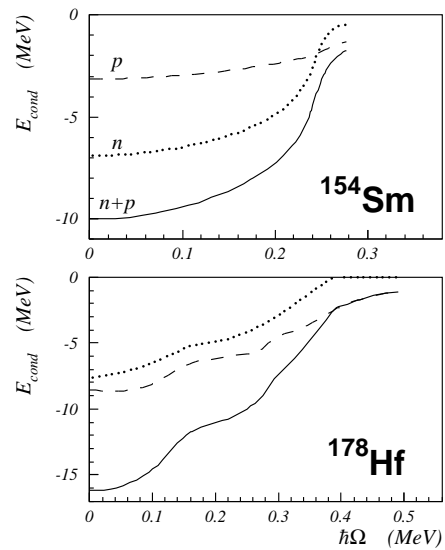}}
  \caption{Condensation energies (in MeV) as functions of the angular velocity
    $\Omega$ (in MeV) for $^{154}$Sm (upper panel) and $^{178}$Hf (lower
    panel). Proton (neutron resp.) condensation energies are represented as
    a dashed line (dotted line resp.) and the total condensation energy is
    represented as a full line. Calculations are performed within the HFB
    formalism.}
  \label{fig:morepairing}
\end{figure}

We have applied the previously described protocol to define the model
parameters $\Omega_c$ and $k$ [see eq. (\ref{eq:omega}) with $\gamma=1$] for 
rotational bands in five nuclei, including the three cases studied in I and
adding the ground-state bands of $^{154}$Sm and $^{178}$Hf.
The Kelvin circulation mean values for protons and neutrons calculated in the
Routhian HFB approach are plotted on Fig. \ref{fig:morekelvin} as a function
of $\Omega$ for these two nuclei. The different trends of these curves for
$^{154}$Sm as opposed to $^{178}$Hf do not reflect the behavior (illustrated on
Fig. \ref{fig:morepairing}) of the condensation energies, but some specific
dynamical properties which will be discussed bellow.

It is generally considered (see, e.g., \cite{PQL02:htda}) that the SIII
\cite{BFG75:SIII} interaction provides extremely good spectroscopic properties
for normally deformed nuclei, while the SkM$^*$ \cite{Bar82:SkM} interaction,
due to its better surface tension, is well suited to the description of
superdeformed or heavy (i.e., with a fissility close to 1) nuclei. Therefore
we have used the former for the calculations of the ground-state bands of
$^{154}$Sm and $^{178}$Hf, and the latter for the three others nuclei
already considered in I.

First, we have obtained values of the critical angular velocities
$\Omega_c$ according to eq. (\ref{omegac}).

For the five nuclei, HFB calculations at zero-spin have been performed to yield
the corresponding pair condensation energies $E_{\mathrm{cond}}(\Omega=0)$.
They have already been shown on Fig. 2 in I
for $^{192}$Hg and $^{254}$No. In the $^{150}$Gd case, we have
``artificially'' constructed a zero-spin solution consistent with the
deformation of the superdeformed band under consideration by constraining the
HFB solution to have the value of the axial mass quadrupole moment of the SD
rotational band states (of course, without constraint one would have obtained
the normally deformed $^{150}$Gd equilibrium solution).
The retained values of $E_{\mathrm{cond}}(\Omega=0)$ are listed in table
\ref{tab:params}.

\begin{figure*}[t]
  \centering
  \includegraphics[scale=1.7]{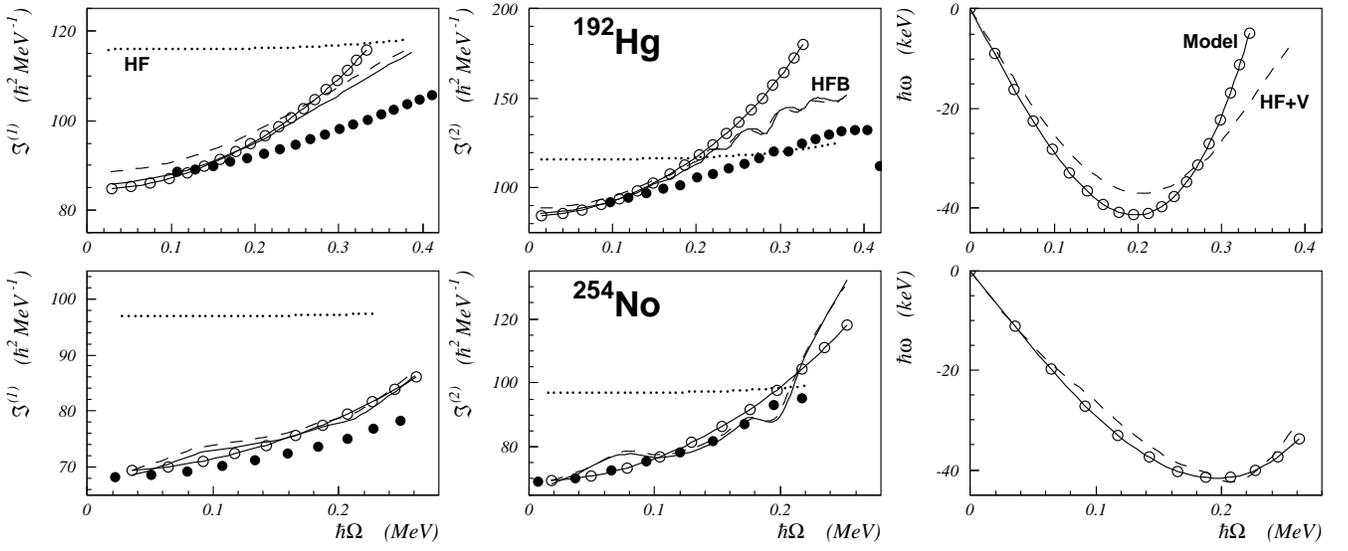}
  \caption{Results of our calculations for $^{192}$Hg (upper panels) and 
  $^{254}$No (lower panels). From left to right we show (as functions of the
  rotation angular velocity $\Omega$ in MeV) for each nuclei the kinetic
  moment of inertia and the dynamical moment of inertia (in
  $\hbar^2\;$MeV$^{\text{-1}}$ units) as well as the Kelvin circulation
  velocity $\omega$ (in keV).  The results obtained within various formalisms
  are represented as follows: dotted line for the (pure) HF formalism, full
  line (dashed line resp.) for the HFB (HF+V resp.) formalisms and full line
  with opened circles for our model. Experimental data for the moments of
  inertia are represented as filled circles. The Lipkin-Nogami correction has
  been applied for the Hg isotope.}
  \label{fig:heavy}
\end{figure*}

\begin{table}[ht]
  \caption{Input parameter $E_{\mathrm{cond}}$ (in MeV),
    $\mathfrak{J}_{\mathrm{ETF}}$ and $\mathfrak{J}_{\mathrm{HFB}}$ (in
    $\hbar^2\:$MeV$^{-1}$ units) for our model (see text). The output parameter
    $\hbar\Omega_c$ (in MeV) and $k$ are also given.}
  \label{tab:params}
  \begin{ruledtabular}
    \begin{tabular}{ce{3}e{3}e{3}e{5}c}
&   \mc{E_{\mathrm{cond}}} & \mc{\mathfrak{J}_{\mathrm{ETF}}} & 
    \mc{\mathfrak{J}_{HFB}} &  \mc{\hbar\Omega_c} & $k$ \\ \hline
$^{150}$Gd & 10.47 &  88.47 &  69.74 & 0.4864 & 0.3443 \\
$^{154}$Sm & 10.03 &  77.85 &  32.23 & 0.5077 & 0.6629 \\
$^{178}$Hf & 16.14 &  97.77 &  26.14 & 0.5745 & 0.6285 \\
$^{192}$Hg &  6.99 & 120.08 &  85.94 & 0.3411 & 0.3159 \\
$^{254}$No &  9.45 & 164.53 &  68.45 & 0.3389 & 0.3196 \\
    \end{tabular}
  \end{ruledtabular}
\end{table}

The values of the semiclassical moment of inertia to be used in eq. 
(\ref{omegac}) (see above the discussion in Section \ref{sec:params}) are also
listed in table \ref{tab:params} together with the resulting $\Omega_c$
values. It is encouraging to note that these values are actually rather close
to the expected values of $\Omega$ which would correspond to a vanishing of
the isoscalar pair correlation energy as exhibited (see Fig. 2 of~I and Fig.
\ref{fig:morekelvin} here).

The values of $A$, $B$, $C$ and $\eta$ have been discussed and given in
Appendix \ref{app:quadratic}. The values of $\mathfrak{J}_{HFB}$ have been
calculated in the close vicinity of $\Omega=0$ and may be deduced from the
plots of $\mathfrak{J}^{(2)}$ of Fig. 1 in I for $^{192}$Hg and $^{254}$No.
Again in the case of $^{150}$Gd this value is yielded by a Routhian HFB
calculation with an appropriate constraint on the axial mass quadrupole
moment (as above discussed). The values of these moments are also listed in
table \ref{tab:params} together with the resulting values of the $k$ parameter.
It is worth noting in Table \ref{tab:params} that the $k$ values corresponding
to a given effective force are not very much varying from one nucleus to
another. However they seem to be strongly dependent of the considered force
(about 0.3 for SkM$^*$ and 0.6 for SIII). We do not yet understand this
feature. 

\begin{figure*}[t]
  \centering
  \includegraphics[scale=1.7]{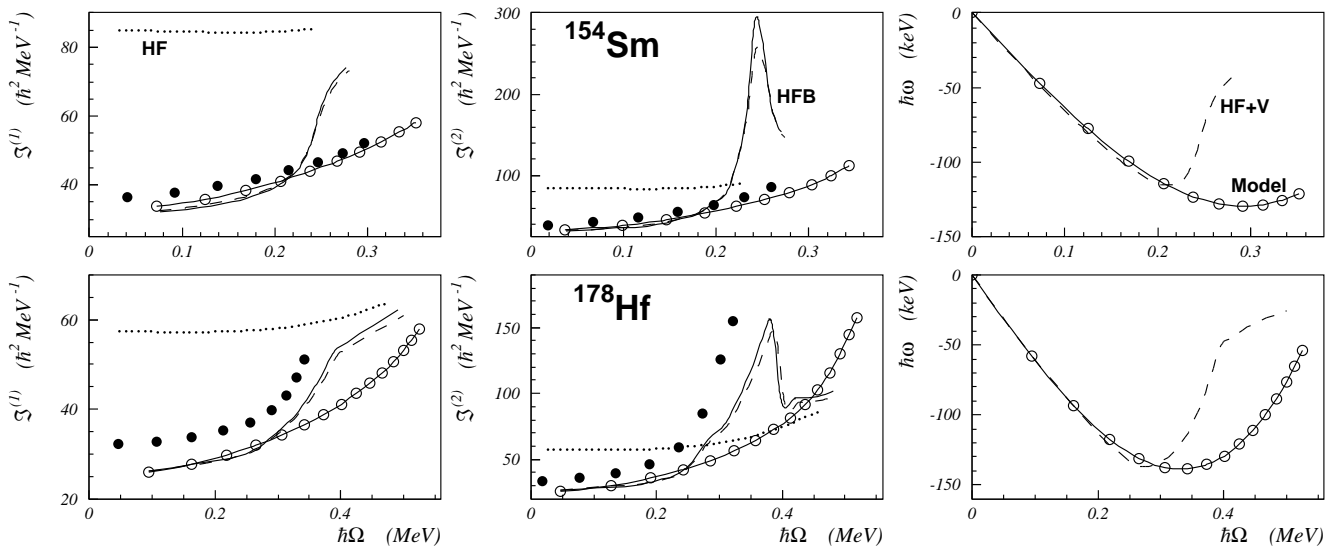}
  \caption{Same as Fig. \ref{fig:heavy} for $^{154}$Sm (upper panels) and 
  $^{178}$Hf (lower panels).}
  \label{fig:light}
\end{figure*}

\begin{figure*}[ht]
  \centering
  \includegraphics[scale=1.7]{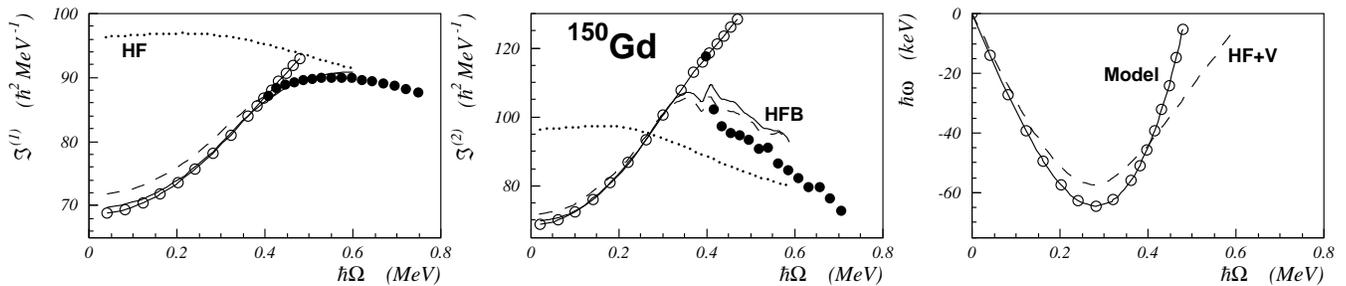}
  \caption{Same as Fig. \ref{fig:heavy} for $^{150}$Gd. The Lipkin-Nogami
    correction has been applied.}
  \label{fig:gado}
\end{figure*}

As a result, the curves $\omega(\Omega)$ yielded by the above defined
parameters are plotted on Figs. \ref{fig:heavy}--\ref{fig:gado} in comparison
with the self-consistently calculated ones (HF+V). The global agreement is
rather satisfactory in a qualitative fashion for $^{150}$Gd, $^{192}$Hg and
$^{254}$No. This is  not so for $^{154}$Sm and $^{178}$Hf where a sudden
raise of $\omega$ occurs. In these three figures, experimental data are taken
from Refs. \cite{TabIso} for $^{154}$Sm and $^{178}$Hf, \cite{Fall91:gd}  for
$^{150}$Gd, \cite{Gall94:hg} for $^{192}$Hg and
\cite{Rei99:no,Lei99:no,Rei00:no} for $^{254}$No.

The final step of the assessment of our model consists of course in performing
Routhian HF+V calculations with the model values of $\omega(\Omega)$ and
compare their results with those of corresponding Routhian HFB calculations.
They should also be compared with the results of Routhian HF+V
calculations where the constrained $\langle{\hat K}_1\rangle$ expectation
values are those obtained in Routhian HFB calculations.
However we knew already from I (see Fig. 1 therein) that both are very close
indeed.

The results of the comparison fall into three categories. For the $^{192}$Hg and
$^{254}$No nuclei (Fig. \ref{fig:heavy}), the agreement between the HFB (and
thus HF+V) and the model results are very good indeed. One exception however
should be noted, it concerns the high spin part of the superdeformed band of
$^{192}$Hg where pairing correlations are strongly damped or disappearing in
the HFB approach and therefore the simple ansatz here is obviously lacking any
sound basis.

The $^{154}$Sm and $^{178}$Hf cases (Fig. \ref{fig:light}) are interesting in
that HFB calculations exhibit a up-bending pattern for both
$\mathfrak{J}^{(1)}$ and $\mathfrak{J}^{(2)}$. The fact that it corresponds
very well to the experimental situation in the latter and is completely absent
in the data for the former is irrelevant for our discussion here. It is
remarkable that our model just ignores these bendings. Therefore we must
conclude that these patterns are not to be attributed to a pairing-rotation
coupling but to something else which is (as well known) the intrusion of
quasiparticle degree of freedom in the rotational dynamics.

As for the last nucleus of our study ($^{150}$Gd), Fig. \ref{fig:gado}
displays clearly in its $\mathfrak{J}^{(1)}$ and $\mathfrak{J}^{(2)}$ parts
that obviously the quasiparticle degrees of freedom are dominating there in
explaining the rotational behavior within this superdeformed band.
Nevertheless, it is to be noted that our model works fairly well in the low
angular velocity regime and only fails above $\hbar\Omega\simeq0.4$ MeV after
the occurrence of a change in the ordering of quasiparticle states.

Let us summarize what has been learned from such a comparison. Actually, one
should conclude at two different levels.

First of all, it appears in some of the studied cases that one could skip
Routhian HFB calculations at finite values of the angular momentum and replace
them by more handy constrained Routhian HF calculations. For that one needs
only to perform a HFB calculation in the close vicinity of $\Omega=0$. This
result is of some practical value in that it allows to predict quickly what
could be the trend of the moments of inertia with respect to the angular
velocity.

However, the limitation of such an approach is clear. It is only valid in
cases where the rotational (global) collective mode does not couple either
with  deformation modes (no rotational stretching, anti-stretching or triaxial
modes, etc.) nor with single-particle degrees of freedom (no back-bending for
instance). It is only able to give an account which turns out to be rather
good quantitavely of the coupling of the pairing degrees of freedom with the
rotational mode.

Taking at face value the rough assumption made in previous papers
\cite{SQM99:period,LSQ03:equiv} that the collective effects of these
correlations could be mocked up by Chandrasekhar \textit{S}-type ellipsoid
velocity fields together with the Mottelson-Valatin picture \cite{MV60} for
the rotational quenching of pairing correlations, our model provides a good
reproduction of fully self-consistent HFB results. It offers thus reasonable
grounds to assume that both conjectures rather accurately describe the
microscopic coupling at work in rigidly deformed purely collective rotational
bands.

\begin{acknowledgments}
Part of this work has been funded through an agreement (\# 12533) between the
BAS (Bulgaria) and the CNRS (France) and another (\# 97-30) between the JINR
(Russia) and the IN2P3/CNRS (France) which are gratefully acknowledged. One
author (P. Q.) would like to thank the Theoretical Division of the LANL for
the hospitality extended to him during his stay at Los Alamos.
\end{acknowledgments}

\appendix
\section{Comparison between correlation and condensation energies}
\label{app:forschle}
Within the HFB formalism, using different interactions in the particle-hole
and particle-particle channel, one cannot obtain a relevant definition of the
pair correlation energy which should be the energy difference between the
correlated (HFB) and uncorrelated (HF) states. On the contrary, the pair
condensation energy which is defined as the ``abnormal'' part of the HFB
energy proportional to $\mathop{Tr}(\kappa\Delta)$, where $\kappa$ is the
``abnormal'' density matrix and $\Delta$ is the pairing tensor, is free of any
ambiguity.

\begin{figure}[h]
  \centering
  \includegraphics[scale=1.7]{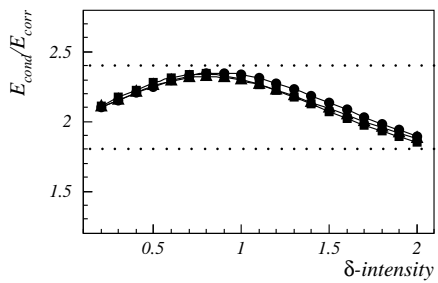}
  \caption{Ratio of the condensation energy with respect to the correlation
    energy as a function of the pairing force intensity (relative to a
    somewhat arbitrary nominal value) for the three nuclear states considered
    in I. The symbol conventions are the same as those in
    Fig.~\ref{fig:kelvin}}.
  \label{fig:condens}
\end{figure}

An alternate formalism to describe pairing correlations in nuclei, dubbed as
``Higher Tamm-Dancoff Approximation'' (HTDA), free of particle number
symmetry-breaking, has been recently presented in Ref. \cite{PQL02:htda}.
Within this formalism, the correlated state is obtained---through the
diagonalisation of a residual interaction in a many-body basis---as a weighted
sum of wavefunctions constructed as particle-hole excitations of a
Slater determinant vacuum. There, it is possible to define consistently a pair
correlation energy as well as a pair condensation energy.
The former is the difference between the HTDA energy and the HF energy of the
vacuum Slater determinant, while the latter is the difference between the HTDA
energy and the HF energy functional associated to the correlated one-body
reduced density. 
Actually, this definition of the pair condensation energy is totally
unequivocal in the case considered bellow where the vacuum is not
self-consistently defined from the correlated $\rho$ matrix (i.e., the non
self-consistent case of Ref. \cite{PQL02:htda}).

We have performed HTDA calculations for the three nuclear states considered 
in I, using a $\delta$-type pairing interaction with intensities varying from
0.2 to 2 times the nominal intensity given in Ref. \cite{Ha}.
The ratio of the pair condensation energy with respect to the pair
correlation energy obtained within this formalism plotted in
Fig. \ref{fig:condens} is seen to change rather slightly around 2 as the
pairing interaction varies.

\section{mass parameters within the HF+V formalism}
\label{app:quadratic}

For the three considered nuclear states we have computed the collective kinetic
energy surface on a 4x4 mesh for angular velocities $\Omega$ and $\omega$
lower in absolute value than 30 keV$\hbar^{-1}$. This energy surface has then been fitted
by the quadratic expression of eq. (\ref{eq:quadratic}). The mass
parameters $A$, $B$ and $C$ so obtained do take into account the
Thouless-Valatin self-consistent contributions and are listed in table
\ref{tab:masses}. The Inglis approximation of these mass parameters as
obtained by eqs.~(\ref{eq:inglis}) are also given.
\begin{table}[h]
  \caption{Mass parameters of the collective energy in HF+V calculations for
    the five considered nuclear states in $\hbar^2\;$MeV$^{-1}$ units.}
\label{tab:masses}
  \begin{ruledtabular}
  \begin{tabular}{le{3}e{3}e{3}e{3}e{3}e{3}}
  &\multicolumn{3}{c}{Inglis}&\multicolumn{3}{c}{Thouless-Valatin}\\
  & \mc{C} & \mc{B} & \mc{A} & \mc{C} & \mc{B} & \mc{A}
  \\ \hline
  \rule{0pt}{3ex}%
  $^{150}$Gd &  88.52 &  74.79 &  86.49 &  96.92 &  81.40 &  94.16 \\
  \rule{0pt}{3ex}%
  $^{154}$Sm &  67.26 &  62.57 &  63.47 &  84.92 &  78.81 &  79.14 \\
  \rule{0pt}{3ex}%
  $^{178}$Hf &  48.17 &  43.37 &  44.24 &  57.40 &  51.38 &  51.88 \\
  \rule{0pt}{3ex}%
  $^{192}$Hg & 106.49 &  92.26 & 101.46 & 115.88 &  99.27 & 108.65 \\
  \rule{0pt}{3ex}%
  $^{254}$No &  86.70 &  78.51 &  81.48 &  96.95 &  87.50 &  89.98
  \end{tabular}
  \end{ruledtabular}
\end{table}

\begin{table}[h]
  \caption{Enhancement of the Thouless-Valatin mass parameters relative to the
    Inglis ones (in percent) within the HF+V formalism.}
  \label{tab:eta}
  \begin{ruledtabular}
  \begin{tabular}{lddd}
  &\mc{C} & \mc{B} & \mc{A} \\ \hline
  \rule{0pt}{3ex} $^{150}$Gd &  9.5~\% &  8.8~\% &  8.9~\% \\
  \rule{0pt}{3ex} $^{154}$Sm & 26.3~\% & 26.0~\% & 24.7~\% \\
  \rule{0pt}{3ex} $^{178}$Hf & 19.2~\% & 18.5~\% & 17.3~\% \\
  \rule{0pt}{3ex} $^{192}$Hg &  8.8~\% &  7.6~\% &  7.1~\% \\
  \rule{0pt}{3ex} $^{254}$No & 11.8~\% & 11.5~\% & 10.4~\% 
  \end{tabular}
  \end{ruledtabular}
\end{table}

As expected, the Thouless-Valatin values are higher than those obtained with
the Inglis approximation. The enhancement (due to the taking into account of
all time-odd self-consistent contributions) is shown in table \ref{tab:eta}.
In view of the values exhibited in this table, we can assume that for the
three considered nuclear states, the Thouless-Valatin correction roughly
enhance the Inglis mass parameters by 10 \% for the SKM$^*$ interaction and 20
\% for SIII.

\section{Moment of inertia within doubly-constrained Routhian calculations}
\label{app:inertia}
The doubly-constrained Routhian in use in HF+V calculations within the present
model writes (see Refs. \cite{MQS97:vort,SQB99:HF,SQM99:period}):
\begin{equation}
  \label{eq:routh}
  \hat R = \hat H - \Omega \hat J_1 - \omega(\Omega) {\hat K}_1 \,.
\end{equation}
The stationary condition $\delta\langle\hat R\rangle=0$, at a fixed value of
$\Omega$, entails
\begin{equation}
  \frac{1}{\Omega} \frac{\mathrm{d} \:\langle\hat H\rangle}{\mathrm{d} \Omega}
  = \frac{\mathrm{d} \:\langle\hat J_1\rangle}{\mathrm{d} \Omega}
  + \frac{\omega(\Omega)}{\Omega}
  \frac{\mathrm{d} \:\langle\hat K_1\rangle}{\mathrm{d} \Omega}\,.
\end{equation}
Within a purely rotational formalism, only the first term of the right hand
side would be present and would be identified with the dynamical moment
of inertia.
The second term which arises here is therefore directly related to the added
intrinsic vortical mode and one is led to the conclusion that the dynamical
moment of inertia in our model should be taken as
\begin{equation}
  \mathfrak{J}^{(2)} = \frac{\mathrm{d} \:\langle\hat J_1\rangle}{\mathrm{d}
    \Omega}\,.
\end{equation}
This is actually what had been assumed in I where it was substantiated by the
fact that requiring identical values of $\langle\hat J_1\rangle$ and
$\langle\hat K_1\rangle$ within both HFB and HF+V formalisms yielded the same
angular velocity $\Omega$ in the two calculations and thus the same moments of
inertia.

\bibliography{simplemodel,perso,history,hf,book,these,vort,scldm,stagg,expe}

\end{document}